\documentclass[twocolumn,prl,superscriptaddress,showpacs]{revtex4}

\usepackage{amssymb}
\usepackage[dvips,final]{graphicx}
\usepackage{dcolumn}
\usepackage{bm}
\begin{document}

\title{Influence of Charge order on the Magnetic Properties of Na$_{x}$CoO$_{2}$
for $x>0.65$}

\author{I.R.~Mukhamedshin}
\affiliation{Laboratoire de Physique des Solides, UMR 8502, Universit\'e
Paris-Sud, 91405 Orsay, France
}%
\affiliation{Physics Department, Kazan State University, 420008 Kazan,
Russia
}%

\author{H.~Alloul}%
\email{alloul@lps.u-psud.fr} \affiliation{Laboratoire de Physique des
Solides, UMR 8502, Universit\'e Paris-Sud, 91405 Orsay, France
}%

\author{G.~Collin}
\affiliation{ Laboratoire L\'eon Brillouin, CE Saclay, CEA-CNRS, 91191
Gif-sur-Yvette, France
}%

\author{N.~Blanchard}%
\affiliation{Laboratoire de Physique des Solides, UMR 8502, Universit\'e
Paris-Sud, 91405 Orsay, France
}%

\begin{abstract}
We have synthesized and characterized the four different stable phases of Na
ordered Na$_{x}$CoO$_{2}$, for $0.65<x\lesssim 0.75$. Above 100K they
display similar Curie-Weiss spin susceptibilities as well as ferromagnetic
$q=0$ spin fluctuations in the CoO$_{2}$ planes revealed respectively by
$^{23}$Na NMR shift and spin lattice $T_{1}$ data. The Co disproportionate
already above 300K into Co$^{3+}$ and\  $\approx $Co$^{3.5+}$ in all phases,
which allows us to understand that magnetism is favoured. Below 100K the
paramagnetic properties become quite distinct, and a 3D magnetic order sets
in only for $x=0.75$, so that charge order has a subtle incidence on the low
$T$ energy scales and transverse magnetic couplings.
\end{abstract}


\maketitle

The lamellar cobaltates display diverse physical properties including high
thermoelectric power, superconductivity, Curie-Weiss \cite{Foo} and ordered
antiferromagnetism (AF) \cite{Sugiyama}, which are controlled for instance
by the Na content inserted between the CoO$_{2}$ layers. Their electronic
properties are expected to be constructed from the low spin states of Co$%
^{3+}$ $(S=0)$ and Co$^{4+}$ $(S=1/2)$ ions in the large crystal field which
prevails in the edge sharing Co oxygen octahedra building the Co triangular
lattice. By analogy with the cuprates, AF is expected for the hypothetical
CoO$_{2}$ half filled Mott-Hubbard insulator and should disappear with
increasing $x$. Unexpectedly, AF phases are found far from this Mott limit,
for $x\geqslant 0.75$ \cite{Mendels}, which rather corresponds to hole
doping of Na$_{1}$CoO$_{2}$. This band insulator is built from the non
magnetic Co$^{3+}$ state \cite{Lang}, for which the $t_{2g}$ multiplet is
filled by the six $d$ electrons. The hole doped AF\ phases are ferromagnetic
in plane and AF between planes (A type AF) \cite{Boothroyd,Bayrakci}. The 3D
dispersion of the spin wave excitations found by Inelastic Neutron
Scattering (INS) has been analyzed with Heisenberg Co-Co AF coupling between
planes either with nearest \cite{Boothroyd,Bayrakci} or next nearest
neighbour exchange through Na orbitals \cite{Mazin}.

While all these approaches assume a homogeneous state for the Co sites, $%
^{23}$Na NMR allowed us to demonstrate that Na atomic order sets in for a
specific $x_{0}\approx 0.7$ phase, together with a disproportionated Co
charge order \cite{NaPaper}. In this metallic Curie Weiss phase, fixed non
magnetic Co$^{3+}$ sites coexist with holes delocalized on magnetic Co sites
with an average valence $\approx $Co$^{3.5+}$ \cite{CoPaper}. Is this
restricted to the Na ordering of this $x_{0}\approx 0.7$ phase? This open
question is all the more important as Na atomic order patterns have been
suggested, or seen \cite{Zandbergen,Roger} for various $x$ values. Are the
singular magnetic properties driven by specific Na orders? Such a scenario
appears likely as a distinct in plane AF order sets in at $T_{N}=$86K  for
the singular  $x=0.5$ phase \cite{Mendels,Gasparovic}, and is followed by a
metal insulator transition at 50K which does not result from the
Co$^{3+}$/Co$^{4+}$ charge differentiation anticipated initially
\cite{Bobroff}.

In order to check the evolution from $x_{0}\approx 0.7$ to the AF phases, we
have synthesized and investigated with $^{23}$Na and $^{59}$Co NMR all the
distinct phases from $x=0.67$ to the first well characterized AF phase. This
constitutes the first systematic effort to study the influence of $x$ and
charge order on the static and dynamic susceptibilities. Our data will be
shown to reveal that the CoO$_{2}$ planes display common magnetic properties
above 100K, but that the lower $T$ properties are due to low energy
modifications of the band structure of the correlated metallic state, which
are apparently governed by the distinct Na orderings of the phases found.

\begin{table}[b]
\caption{Parameters of the studied phases.The accuracy on their difference
of Na content $x$ is much better than that on $x$ itself ($\pm 0.01$); $c$
axis ($\pm 2\cdot 10^{-4}$\AA ); real cell = superlattice or incommensurate
modulation q(b*) determined by X ray diffraction; Co$^{3+}$ fraction $y$
obtained from $^{59}$Co NMR.}
\label{table1}%
\begin{ruledtabular}
\begin{tabular}{cccccc}
Phase & $x$ & $c$ (\AA) & real cell & y (\%) Co$^{3+}$\\
\hline
H67 & $\approx .67$ & 10.938 & $a\surd 3,3a,3c$ & 26(4)\\
O71 & $.70 \leftrightarrow .71$ & 10.900-.888 & 0.28$\leq$ q(b$^{*}$)$\leq$ 0.29 & 40(5)\\
H72 & $\approx .715$ & 10.879 & q(b$^{*}$)$\approx$0.281 & 37(5) \\
H75 & $\geq .75$ & 10.807 & not determined & 33(4) \\
\end{tabular}
\end{ruledtabular}
\end{table}

Using $^{23}$Na and $^{59}$Co NMR spectra combined with Rietveld refinements
of x ray data we have evidenced that multiphasing is quite common in Na$_{x}$%
CoO$_{2}$ samples. We could sort out four different stable phases for $%
0.65\leqslant x\lesssim 0.75$, and synthesized reproducibly nearly pure
samples, which requires a continuous control and no exposure to air. These
phases all have two layer Co structures with the hexagonal (P6$_{3}$/mmc)
unit cell as reference substructure lattice. However, as summarized in Table~%
\ref{table1} the homogeneity range is sequenced in four distinct narrow $x$
domains, each with a specific Na ordering leading to characteristic
additional diffraction: incommensurate satellites or superstructure
commensurate reflexions. These orderings result in symmetry lowering and the
reference subcell is systematically orthorhombic (Ccmm, n$^{\circ }$63): $%
a_{ort}=a_{hex}\sqrt{3}$; $b_{ort}=a_{hex}$; $c_{ort}=c_{hex}$. Among these
phases a single one, $x=0.71$, exhibits a significant distortion with
respect to the hexagonal substructure, with $a_{ort}/\sqrt{3}\approx 2.84$%
\AA , $b_{ort}\approx 2.83$\AA , hence the labelling H67, O71, H72 and H75
used hereafter and in Table~\ref{table1}. The formerly studied $x_{0}\approx
0.7$ phase \cite{NaPaper,CoPaper} has in fact the lowest $x$ value $%
x=0.67(1) $ reached for any material when it evolves in insufficiently dry
atmosphere.

\begin{figure}[tb]
\center
\includegraphics[width=1\linewidth]{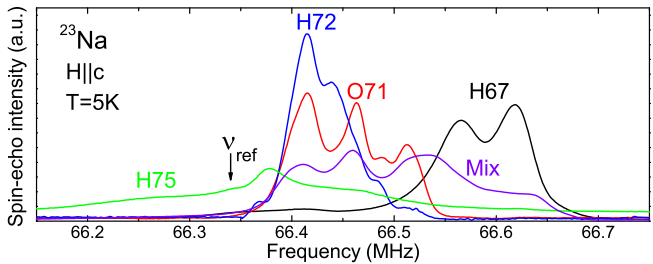}
\caption{(color online) $^{23}$Na NMR\ central line spectra taken at 5K.
They are quite distinct for the four nearly pure phases, with some overlap
between O71 and H72 spectra. That for a sample which is clearly a mixture of
H67, O71 and H72 is shown as well. $\protect\nu _{ref}$. is the non magnetic
$^{23}$Na NMR reference.}
\label{fig:1}
\end{figure}

The single crystal grains of these samples were oriented in the $H_{0}=7$%
~Tesla NMR field within Stycast or paraffin. As reported in Ref.%
\onlinecite
{NaPaper} on H67, the $^{23}$Na NMR displays quadrupole splittings which
allow us to distinguish the different Na sites in the structure. In view of
the more complex structures of the new phases there was no surprise in
finding more resolved sites, with splittings quite similar in magnitude. The
magnetic properties of the compounds are probed through the NMR shifts of
the different Na sites resolved in the $(-\frac{1}{2}\leftrightarrow \frac{1%
}{2})$ transition of the $^{23}$Na spectra presented in Fig.\ref{fig:1}.
There one can see that the 5K \ spectra are quite distinct for the four
phases. For H75 a large broadening occurs in the AF state below $T_{N}=22$K,
while the spectrum of H67 is much more shifted than those of O71 and H72,
which are distinct but display some overlap. As the H72 batch has been found
to evolve fast at room $T$ towards O71, a slight mixture of the two pure
phases might be unavoidable in these samples. Quite generally $^{23}$Na NMR
allowed us to control phase purity, as multiphase samples display
superimposed spectra, as seen in Fig.\ref{fig:1}.

\begin{figure}[tb]
\center
\includegraphics[width=1\linewidth]{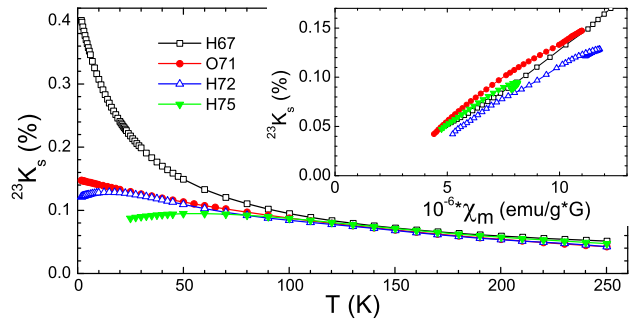}
\caption{(color online) $T$ dependence of the mean $^{23}$Na NMR shift.
Identical behaviour above 100K can be seen for the four phases with
remarkable differences at low $T$. Inset: the linear variations of $K_{s}$
versus SQUID data for $\protect\chi _{m}$ underlines the purity of the
samples. The H67 data for $T<$30K \protect\cite{NaPaper} has been omitted to
better display those of the new phases.}
\label{fig:2}
\end{figure}

Let us recall, as detailed in Ref.\onlinecite {NaPaper} that for a field $%
H_{0}\parallel \alpha $, the NMR shift $K_{\beta }^{\alpha }$ of a Na atomic
site $\beta $ probes the spin susceptibility $\chi _{s,i}^{\alpha }(T)$ of
the neighbouring Co sites $i$ through transferred hyperfine couplings $%
A_{\beta ,i}^{\alpha }$ with $K_{\beta }^{\alpha }=\sum_{i}A_{\beta
,i}^{\alpha }\;\chi _{s,i}^{\alpha }(T)$. The main result found for H67, and
verified here for the other phases, is that the $K_{\beta }^{\alpha }(T)\;$%
variations\ scale with each other for all Na sites. This $T$ dependence is
associated with the average $\chi _{s}^{\alpha }(T)$ of the magnetic Co
sites of the structure. So, overlooking the diversity of Na sites, the first
moment (or center of gravity) of the $^{23}$Na NMR signal writes $%
K_{s}^{\alpha }=A_{eff}^{\alpha }\chi _{s}^{\alpha }(T)$ where $%
A_{eff}^{\alpha }$ is an effective hyperfine field per Co site. The $T$
variations of $K_{s}^{\alpha }$ are reported in Fig.\ref{fig:2}, and are
shown to be quite identical for $T>100$K with a unique Curie-Weiss $%
(T+\Theta )^{-1}$ variation (with $\Theta \approx $80K). They surprisingly
differ markedly below 100K, the low $T$ enhancement of $\chi _{s}^{\alpha
}(T)$ observed for H67 being progressively reduced for increasing $x$.

In the H75 AF phase, the saturation of $K_{s}(T)$, that is $\chi
_{s}^{\alpha }(T)$, seen at low $T$ in Fig.\ref{fig:2} should be associated
with the onset of AF correlations. In a uniform Heisenberg model, one would
then assign the progressive increase of $\chi _{s}^{\alpha }(T)$ at low $T$
with decreasing $x$ to a decrease of $T_{N}$ and of out of plane AF coupling
strength. However this primary interpretation fails as NMR data taken down
to 1.4K (and $\mu $SR to 50mK \cite{Mendels2}), did not evidence any frozen
magnetic state in the three other phases, which are then \textit{paramagnets
in their ground state}, most probably metallic, as no electronic transition
is detected.

SQUID measurements of the macroscopic susceptibility $\chi _{m}$ taken in
5~T also revealed the different magnetic properties of the 4 phases. The
slopes of the $K_{s}$ versus $\chi _{m}\;$plots reported in Fig.\ref{fig:2}
yield similar values for the hyperfine coupling ($A_{hf}^{c}$= 9.1(3),
8.0(3), 7.3(3) and 7.8(3)KGauss/$\mu _{B}$ from H67 to H75 respectively),
which could be expected as $^{23}$Na sites are coupled with many Co \cite
{NaPaper}. In all phases the anisotropy of $\chi _{s}^{\alpha }(T)$, given
by that of $K_{s}^{\alpha }$, does not exceed $\pm $0.1.

\begin{figure}[tb]
\center
\includegraphics[width=1\linewidth]{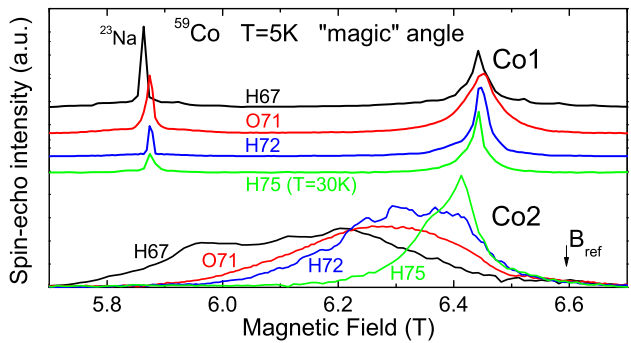}
\caption{(color online) Spectra taken with a large pulse spacing $\tau
=200~\protect\mu s$ in a spin echo sequence allowed us to isolate the narrow
spectra of Co1 sites with long $T_{2}$. The broad spectra of the magnetic
Co$_{m}$ with short $T_{2}$ are obtained by subtracting the Co1 spectra from
those taken with $\protect\tau =10~\protect\mu s$. The Co$_{m}$ shifts
decrease with $x$ as the $^{23}$Na shift does in Fig.\ref{fig:1} and
\ref{fig:2}.} \label{fig:3}
\end{figure}

The difference between phases was also clearly seen in the $^{59}$Co NMR
spectra, which also display many Co sites as for H67 \cite{CoPaper}. We
identified two classes of Co sites, their $(-\frac{1}{2}\rightarrow
\frac{1}{2})$ transitions being better visualized in Fig.\ref{fig:3} when
$H_{0}$ is at the ''magic angle'' for which quadrupole effects are reduced
(54.7$^{\circ }$ from the $c$ axis). A first series, the Co1 ''class'', are
non magnetic Co$^{3+}$ with small spin lattice $T_{1}^{-1}$ and spin spin
$T_{2}^{-1}$ relaxation rates and similarly small NMR shifts $\approx 2\%$
in the four phases, as compared to the magnetic sites to be discussed below.
Their shift is dominated by the $T$ independent isotropic orbital
susceptibility of the Co$^{3+}$ ion (1.95\% in insulating Na$_{1}$CoO$_{2}$
\cite{Lang}). The complex spectra of the fast relaxing magnetic Co$_{m}$
''class'', which involved Co2 and Co3 sites in H67 \cite{CoPaper}, are also
shown in Fig.\ref{fig:3}. They include diverse unresolved sites with an
average NMR shift increasing markedly with decreasing $x$ in perfect
agreement with expectations from $^{23}$Na NMR data.

The fraction $y$ of Co$^{3+}$ sites, estimated from the Co1 relative NMR
intensity (corrected for $T_{2}$ decay), increases slightly, but not
regularly with $x$ (Table~\ref{table1}), this overall trend being expected
as all Co sites become Co$^{3+}$ for $x=1$. However as $y<1-x$, the average
valence of the Co$_{m}$ sites is always much smaller than Co$^{4+}$, and the
charge disproportionation detected in H67 is present in all phases,
\textit{including the AF ordered phase}.

Our data involves various evidences that this charge distinction occurs
already above room $T$, contrary to the proposal of
Ref.\onlinecite{Gavilano}, as i) we could detect the Co1 sites up to room
$T$, and the Co$_{m}$ up to 220K, at least in H67, ii) the perfect scaling
of $^{23}$K with $\chi _{m}$ in Fig.\ref{fig:2} applies up to room $T$, well
above the onset of Na motion at $\approx $200K detected hereafter from
$^{23}$Na $T_{1}$ data. These findings, and more refined observations
\cite{Foot1} prove as well that the Co charge disproportionation is
correlated with the Na environment (e.g. Na1 sites being on top of
Co$^{3+}$).

\begin{figure}[tb]
\center
\includegraphics[width=1\linewidth]{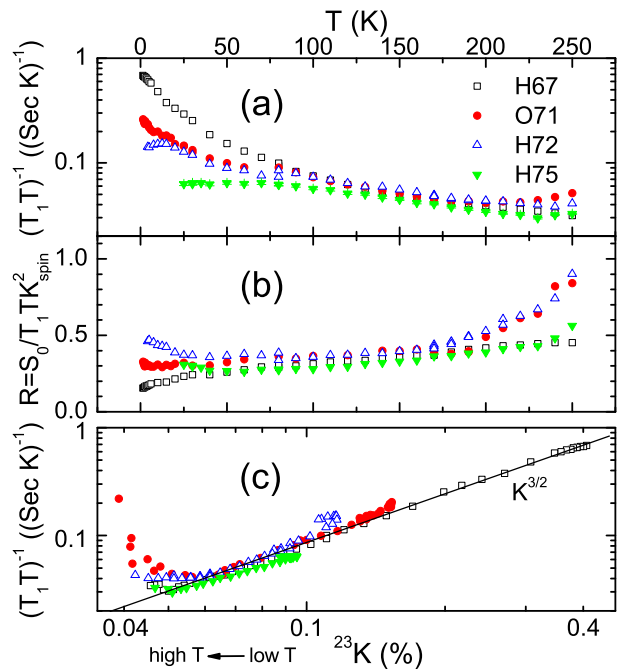}
\caption{(color online) $T$ variation of $(T_{1}T)^{-1}$ (a) and the
normalized Korringa product $R$ (b) for the four phases. While the data are
distinct below 100K, in (c) a universal scaling between $(T_{1}T)^{-1}$ and
$^{23}K$ is shown to apply. The high $T$ deviations due to Na motion and the
slight low $T$ increases for H72 and O71 are discussed in the text.}
\label{fig:4}
\end{figure}

To search for differences in the dynamic electronic susceptibilities of
these phases we have taken extensive $^{23}$Na $T_{1}$ data. As $^{23}$Na
has a spin $I=3/2$, its nuclear magnetization recovery should be given by
\[
M(t)\propto M_{0}(1-W\exp (-6t/T_{1})-(1-W)\exp (-t/T_{1})),
\]
with $W=0.9$ if only the central transition has been saturated. This
condition being impossible to fulfill strictly experimentally, $W$ has been
left as an adjustable parameter, which was found to evolve between 0.9 and
0.7 depending on the sample and experimental conditions. The $T_{1}^{-1}$
data were found slightly anisotropic, i.e. $\approx $30\% larger for $%
H_{0}\perp c$ than for $H_{0}\parallel c$, for $T<$200K. So, in Fig.\ref
{fig:4} we only plotted the data for $H_{0}\parallel c$.

As might be expected, all phases display nearly the same $(T_{1}T)^{-1}$
above 100K, $^{23}K(T)$ being identical as well (Fig.\ref{fig:1}). These two
quantities can be correlated in the Korringa ratio $%
R=S_{0}/T_{1}T(^{23}K)^{2}$, which measures the deviation with respect to
the Fermi liquid value $S_{0}=(\hbar /4\pi k_{B})(\gamma _{e}/\gamma
_{n})^{2}$. In Fig.\ref{fig:4}b $R$ is indeed found sample independent
between 80 and 160K. An extra high $T$ contribution to $T_{1}^{-1}$ is
assigned to fluctuations induced by Na motion, their onset taking place at
distinct $T$ for the different phases. Below 160K, $R$ is always smaller
than unity, as has been noticed in a $x\approx 0.7$ mixed phase sample by
Ihara \emph{et al.} \cite{Ishida}. As recalled there, $R<1$ is an evidence
that the excitation spectrum is dominated by ferromagnetic $\mathbf{
q}\approx 0$ fluctuations that enhance markedly $\chi _{s}(\mathbf{q}=0),$
that is $^{23}K_{s}$, while $\left( T_{1}T\right) ^{-1}$ is less enhanced as
it probes $\chi ^{\prime \prime }(\mathbf{q},\omega )$ at all $\mathbf{q}$
values. This corroborates the observation  by INS \cite{Boothroyd} of  a
quasielastic peak at $\mathbf{q}\approx 0$ above $T_{N}$ in the H75 phase.
The identical behaviour found here above 100K for all samples extends this
result to all our Curie-Weiss paramagnetic phases.

Below 100K the data for $\left(T_{1}T\right)^{-1}$ and $R$ reported in
Fig.\ref{fig:4} differ markedly for the various phases, which reflects the
distinct ground states revealed by the shift data in Fig.\ref{fig:2}. To
sort out whether the fluctuation spectrum is modified at low $T$ we searched
for relationships between $(T_{1}T)^{-1}$ and $^{23}K$, as done in Fig.\ref
{fig:4}c. Spin fluctuation theories in nearly ferromagnetic metals are known
to give $(T_{1}T)^{-1}=aK^{n}$ \cite{Moriya-Ueda}, with $n=1$ in 3D \cite
{Alloul-Mihaly}, while $n=3/2$ is expected for 2D \cite{Hatatani-Moriya}.
For H67, we do remarkably find an accurate scaling with $n=1.5\pm 0.1$ over
the entire range 1.5K$<T<$300K. Of course, $^{23}K$ and $(T_{1}T)^{-1}$
being identical above 80K for all phases, apart the Na motion contribution,
this scaling applies as well. Ferromagnetic 2D fluctuations are indeed
characteristic of all phases at high $T$. But, the very impressive
experimental result in Fig.\ref{fig:4} is that the $\left( T_{1}T\right)
^{-1}=aK^{3/2}$ scaling law is seen to extend down to low $T$ with a unique
$a$ coefficient for all phases \cite{Foot3}, whatever the behaviour of $\chi
_{s}(\mathbf{q}=0)$, including the AF phase down to $T_{N}$.

At this stage one might wonder whether the saturation of $^{23}K(T)$
observed below 100K in H75 is related to AF plane to plane couplings. Such
AF fluctuations that enhance $\chi ^{\prime \prime }(\mathbf{q}=\mathbf{q}%
_{AF})$ should result in an increase of $(T_{1}T)^{-1}$ and $R$ with a
divergence at $T_{N}$. But they are not probed by the $^{23}$Na nuclei, as
the local fields induced by two adjacent Co layers cancel in the A type AF
structure, as confirmed by the weak $^{23}$Na NMR shift in the N\'{e}el
state (Fig.\ref{fig:1}). The $^{23}$Na $T_{1}$ only probes then the strength
of the ferro fluctuations, and the perfect scaling found between $%
(T_{1}T)^{-1}$ and $^{23}K^{3/2}$ is then a proof that the main incidence of
AF fluctuations is to reduce the ferromagnetic ones below 100K. For H67, the
comparison with the numerical results of Hatatani \emph{et al.} \cite
{Hatatani-Moriya} allows us to point out that the low $T$ increase of $\chi
(\mathbf{q}=0)$ with respect to the common Curie-Weiss variation is exactly
the situation expected by these authors in the immediate vicinity of a
ferromagnetic instability. Therefore the H67 phase \textit{appears as an
ideal 2D nearly ferromagnetic metal without 3D ordering settling in at low}
$T$.

The four phases differentiated in this thorough investigation exhibit
specific Na orderings, but similar Co charge disproportionation, and quasi
identical ferromagnetic in plane fluctuations above 100K which appear
independent of the detailed distribution of Co charges. However, it seems to
us that the occurrence of non magnetic Co$^{3+}$ is essential as it reduces
the number of hopping paths between magnetic sites with respect to a
homogeneous structure. The associated decrease of bandwidth $W$ and increase
of hole density on the remaining magnetic sites magnify the importance of
correlations $U$, which might enhance the ferromagnetic tendency supported
by LDA \cite{Singh} for the homogeneous case.

As for the ground state properties, one would expect AF to be driven by
specific Na orderings in some phases, which would presumably give a hectic
evolution versus $x$, contrary to the smooth one found for $x<$0.75, and to
the abrupt occurrence of AF for any $x>$0.75. In any case metallic magnetism
is the key point in these phases, which prohibits the use of local moment
Heisenberg models. One might consider \cite{McKenzie} that a Fermi liquid
state is only reached below an energy scale given by the temperature
$T^{\ast }$ at which $\chi _{s}$ saturates, which increases from $\approx $
1K for H67 \cite{NaPaper} to $\approx 60$K for H75. The band parameters
associated with Na and Co charge orders would then be responsible for these
$T^{\ast }$ values and for the transverse couplings which drive spin density
wave order. While some attempts have been done to consider both correlations
and charge order \cite{Marianetti}, extensions to diverse local orders are
certainly required to fully explain the evolution with $x$ of the ground
state properties of these materials.

We thank J.~Bobroff, F.~Bert and P.~Mendels for performing the $\mu $SR
measurements on the paramagnetic phases and for helpful discussions, as well
as G.Kotliar, I.~Mazin, F.~Rullier-Albenque and D.~Singh for their
stimulating interest. We acknowledge financial support from the ANR
(NT05-441913) and INTAS (04-83-3891).


\begin{references}
\bibitem{Foot1}  Slight increases of the Co1 NMR shift above the onset $T$ of the Na
motion, prove that it induces an exchange of Co1 and
Co$_{m}$ sites at a slow rate. Details will be given in a full report.
\bibitem{Foot3} For H72  and O71, the small deviations for $T\lesssim 5$K are not intrinsic
 as data for the resolved Na lines were not found to scale perfectly. Purification of these phases is required for accurate low $T$ studies.
\end{references}
\end{document}